# Waveguide single-photon detectors for integrated quantum photonic circuits

J.P. Sprengers[1*], A. Gaggero[2,*], D. Sahin[1], S. Jahanmiri Nejad[1], F. Mattioli[2], R. Leoni[2], J. Beetz[3], M. Lermer[3], M. Kamp[3], S. Höfling[3], R. Sanjines[4] and A. Fiore[1,a]

[1] COBRA Research Institute, Eindhoven University of Technology, PO Box 513, 5600 MB Eindhoven, The Netherlands

[2] Istituto di Fotonica e Nanotecnologie, CNR, Via Cineto Romano 42, 00156 Roma, Italy

[3] Technische Physik, Physikalisches Institut and Wilhelm Conrad Röntgen Research Center for Complex Material Systems, Universität Würzburg, Am Hubland, D-97074 Würzburg, Germany

[4] Institute of Condensed Matter Physics, Ecole Polytechnique Fédérale de Lausanne (EPFL), Station 3, CH-1015 Lausanne, Switzerland

[a] Electronic mail: a.fiore@tue.nl

*These authors contributed equally.

**The generation, manipulation and detection of quantum bits (qubits) encoded on single photons is at the heart of quantum communication and optical quantum information processing[1]. The combination of single-photon sources, passive optical circuits and single-photon detectors enables quantum repeaters[2] and qubit amplifiers[3], and also forms the basis of all-optical quantum gates[4] and of linear-optics quantum computing[5]. However, the monolithic integration of sources, waveguides and detectors on the same chip, as needed for scaling to meaningful number of qubits, is very challenging, and previous work on quantum photonic circuits has used external sources and detectors[6,7,8,9]. Here we propose an approach to a fully-integrated quantum photonic circuit on a semiconductor chip, and demonstrate a key component of such circuit, a waveguide single-photon detector. Our detectors, based on superconducting nanowires on GaAs ridge waveguides, provide high efficiency ($\approx$20%) at telecom wavelengths, high timing accuracy ($\approx$60 ps), response time in the ns range, and are fully compatible with the integration of single-photon sources, passive networks and modulators.**

A linear-optics quantum computer[5] requires hundreds to thousands of single-photon components including sources, detectors and interferometers, which is obviously only feasible in an integrated circuit. Even the small-scale circuits needed in quantum repeaters[2] would greatly benefit from monolithic integration in view of the improved stability and coupling efficiency attainable in a chip. A very large experimental research activity has been dedicated to the development of single-photon sources based on III-V semiconductors[10], in view of large-scale integration, and to passive quantum circuits based on silica-on-silicon[6] and on laser-micromachined glass[8,9], but a clear approach towards a fully integrated photonic network including sources and detectors has not been proposed. This is in large part due to the complexity of most single-photon detector technologies – for example, the complex device structures associated to avalanche photodiodes are not easily compatible with the integration of low-loss waveguides and even less of sources. Transition-edge sensors may be suited for integration[11], but they are plagued by very slow response times (leading to maximum counting rates in the tens of kHz range) and require cooling down to <100 mK temperatures. Here we propose a platform for the full integration of quantum photonic components on the same chip. It is based on the mature III-V semiconductor technology and comprises (Fig. 1(a)) waveguide single-photon sources based on InAs quantum dots (QDs), GaAs/AlGaAs ridge waveguides, Mach-Zehnder interferometers using directional couplers or multimode-interference couplers, and waveguide detectors based on superconducting nanowires. Efficient single-photon emission from QDs in a waveguide can be obtained by using photonic crystals (PhCs), e.g. in a cavity side-coupled to a waveguide[12] or using the slow-light regime in PhC waveguides[13], and the photons can then be transferred to ridge waveguides using tapers. Photons emitted by distinct QDs can be made indistinguishable by using electric fields to control the exciton energy[14]. The high index contrast available in the GaAs/AlGaAs system allows circuits with short bending radii, therefore more compact than in the silica platform[6], while the large electro-optic coefficient of GaAs enables compact modulators operating at GHz frequencies. In this letter we report the key missing component, a single-photon detector integrated with GaAs waveguides. Our waveguide single-photon detectors (WSPDs) are based on the principle of photon-induced hot-spot creation in ultranarrow superconducting NbN wires, which is also used in nanowire superconducting single-photon detectors[15] (SSPDs) and can provide ultrahigh sensitivity at telecommunication wavelengths, high counting rates, broad spectral response and high temporal resolution due to low jitter values. In our design (see Fig. 1(b)), the wires are deposited and patterned on top of a GaAs ridge waveguide, in order to sense the evanescent field on the surface. Four NbN nanowires (4 nm-thick, 100 nm wide and spaced by 150 nm) are placed on top of a GaAs (300 nm)/$Al_{0.7}Ga_{0.3}As$ waveguide, and a 1.85 μm-wide, 250 nm-deep ridge is etched to provide 2D confinement. The dimensions of the waveguide have been optimized to obtain maximum absorption by the NbN wires while leaving a 0.5 μm alignment margin between the wires and the side of the ridge. The electric field amplitude and polarization for the fundamental mode, calculated using a finite-element mode solver, is shown in Fig. 1(b) for λ=1300 nm. For this quasi-transverse-electric (TE) mode we calculate a modal absorption coefficient of $\alpha_{abs}$ = 451 cm$^{-1}$ (assuming a refractive index of 5.23 -5.82i [16] for NbN), corresponding to 90% (99%) absorptance after 51 μm (102 μm) propagation length. This very high and broadband absorptance in an ultrathin wire is unique to waveguide geometries, which allow an interaction length limited only by extrinsic waveguide losses. We note that this design with a TE-polarised, tightly-confined mode is optimized for on-chip applications with integrated single-photon sources (quantum dots in waveguides), which emit in the TE polarisation. Indeed, we recently observed TE-polarised single-photon emission from InAs QDs in a PhC waveguide based on a very similar GaAs/AlGaAs heterostructure[17]. In contrast, TM-polarised modes have a complex spatial profile in this waveguide, which makes the fiber coupling inefficient. By increasing the waveguide thickness by 50 nm, well-confined TM modes with high modal absorption coefficients >500 cm$^{-1}$ can be obtained. The absorptance of both TE and TM light would then approach 100%, resulting in a polarisation-independent detection efficiency.

Nanowire WSPDs with the structure depicted in Fig. 1(b) were fabricated on top of a GaAs/AlGaAs ridge waveguide using an optimised procedure for dc magnetron sputtering deposition of ultrathin NbN on GaAs and several steps of electron-beam lithography and reactive-ion etching, see the Methods section. Fig. 2 shows (a) the WSPD contact structure; (b) a detail of the NbN wires after the wire etching step; (c) the ridge waveguide lithography realigned to the wires; and (d) an enlarged view of (c) showing a realignment accuracy better than 100 nm. A 1 mm-long passive access waveguide was used between the nanowire active section with NbN nanowires and the cleaved facet. We note that the nanowires can be placed anywhere on the chip, independently of the epitaxial structure in the waveguide, which provides maximum flexibility in defining the passive and active areas of the quantum photonic circuit.

The WSPDs were characterized in a cryostat at T≈4 K by end-fire coupling light from a continuous wave 1300 nm diode laser into the waveguide through a polarization-maintaining lensed fiber (see Methods). The inset of Fig. 3 displays the current-voltage characteristic measured for a 50 μm-long WSPD, showing a critical current ($I_c$) of 16.9 μA. When sending 1300 nm photons through the fiber, electrical pulses were measured (inset of Figure 4), showing a pulse duration (full-width-half-maximum) of 3.2 ns and a 1/e decay time of 3.6 ns, which corresponds very well to the expected time constant $\tau=L_{kin}/R=3.6$ ns, where $L_{kin}=180$ nH is the wire kinetic inductance (as calculated from the kinetic inductance per square measured in meanders made of similar NbN wires, $L_{\square}=90$ pH/□) and R=50 Ω is the load resistance). Considering that it takes a time ≈3τ to recover 95% of the bias current after detection, we estimate a maximum counting rate close to 100 MHz. The detector count rate was observed to be extremely sensitive to the fiber-waveguide alignment and to their distance, confirming that the detector responds to guided photons and not to stray light propagating along the surface or in the substrate. The count rate was measured to be proportional to the laser power (Fig. 3), proving operation in the single-photon regime. By illuminating the device with a pulsed diode laser, a total jitter of 73 ps was measured on the WSPD output pulse, corresponding to a 61 ps intrinsic detector jitter after correcting for the 40 ps jitter from the laser pulse.

The measured quantum efficiency (QE) (1300 nm, TE polarization) is plotted in Fig. 4 (left axis) as a function of the normalised bias current $I_b/I_c$. The system quantum efficiency (SQE), defined as the number of counts divided by the average photon number in the fiber at the input of the cryostat, reaches 3.4% for a 50 μm-long device. The device quantum efficiency (DQE), defined with respect to the number of photons coupled to the waveguide, was determined by measuring the coupling efficiency in test waveguides (see Methods), and reaches 19.7%. This value is still lower than the calculated absorptance (90% in the 50 μm-long WSPD), which we attribute to a limited internal quantum efficiency (detection probability upon absorption of a photon), and further improvements of film quality and wire etching process may result in notably improved values.

The dark count rate was measured in another cryostat without optical windows at 4.2 K and is presented on Fig. 4 (right axis), showing the usual exponential dependence as a function of the bias current. Much lower dark count rates can be obtained by cooling the device down to 2 K [18].

In conclusion, we have demonstrated integrated waveguide single-photon detectors based on superconducting nanowires on GaAs ridge waveguides. They provide system (device) quantum efficiencies of 3.4% (≈20 %) at 1300 nm, a timing resolution ≈60ps and dead times of few ns. Further optimisation of film deposition and device fabrication may result in efficiencies approaching 100% due to the high absorptance allowed by the waveguide geometry. Higher system QE and polarisation-independence can be obtained by a waveguide design providing a more extended and symmetric mode profile, and by integrating a tapered coupler[19]. Integrated photon-correlation devices and photon-number-resolving detectors[20] are straightforward to realize by integrating several wires on the same waveguide. Furthermore, this technology is fully compatible with the fabrication of quantum photonic circuits on GaAs waveguides, including fast electro-optic modulators, and with single-photon sources based on InAs quantum dots in waveguides, and therefore opens the way to fully integrated quantum photonic circuits including sources and detectors.

Acknowledgements: We acknowledge interesting discussions with D. Bitauld, M. Thompson and J.L. O'Brien. This work was supported by the European Commission through FP7 projects QUANTIP (Contract No. 244026) and Q-ESSENCE (Contract No. 248095) and by Dutch Technology Foundation STW, applied science division of NWO, the Technology Program of the Ministry of Economic Affairs.

**Methods:**

*Nanofabrication*

Nanowire WSPDs were fabricated on top of a GaAs (300 nm)/Al$_{0.75}$Ga$_{0.25}$As (1.5 µm) heterostructure grown by molecular beam epitaxy on an undoped GaAs (001) substrate. A 4.3 nm-thick NbN layer was deposited by dc reactive magnetron sputtering of a Nb target in a N$_2$/Ar plasma at 350 °C, with deposition parameters optimized for GaAs substrates[21], resulting in a critical temperature $T_c$=10.0 K, and a transition width $\Delta T_c$=650 mK. WSPDs were then fabricated using four steps of direct-writing electron beam lithography (EBL), using a high resolution Vistec EBPG 5HR system equipped with a field emission gun with an acceleration voltage of 100 kV. In the first step Ti(10nm)/Au(60nm) contact pads (patterned as a 50 Ω coplanar transmission line) and alignment markers are defined by lift-off using a PMMA mask (Fig 2a). In the second step, the meander pattern is defined on a 180 nm thick hydrogen silsesquioxane (HSQ) mask and then transferred to the NbN film with a (CHF$_3$+SF$_6$+Ar) reactive ion etching (RIE). Fig 2b) shows a scanning electron microscopy (SEM) image of an etched wire. The meandered NbN nanowire (100nm width, 250nm pitch, length 30-100 µm length), still covered with the HSQ mask, is very regular with a width uniformity of about 5%. In the third step, an HSQ-mask for the waveguide patterning is defined by carefully realigning this layer with the previous one. This layer also protects the Ti/Au pads against the subsequent reactive etching process. Fig. 2c) shows an atomic force microscopy (AFM) image of a fabricated SSPD showing realignment accuracy better than 100nm (Fig 2d). Successively, 250nm of the underlying GaAs layer is etched by a Cl$_2$+Ar ECR (electron cyclotron resonance) RIE. Finally, in order to allow the electrical wiring to the TiAu pads, vias through the remaining HSQ-mask are opened using a PMMA mask and RIE in a CHF$_3$ plasma. The waveguides were cleaved leaving a 1 mm-long passive ridge waveguide between the cleaved facet and the WSPD.

*Characterisation*

The WSPDs were tested by end-fire coupling light from a polarization-maintaining lensed fiber into the waveguides mounted on the cold finger of a continuous flow helium cryostat. The fiber is mounted on piezoelectric positioners, allowing the optimization of the coupling, while the electric signal from the detector is collected via a microwave probe mounted on another piezo tower. This piezo tower is thermally anchored to the cold plate to minimize the thermal load to the detector, resulting in an operating temperature

<4K, as estimated from the measured critical currents. The spot size produced by the lensed fiber has a nominal diameter of 2.5±0.5 μm with a working distance of 14±2 μm. An external optical microscope allows imaging of the sample through the top cryostat window.

For determining the number of photons coupled into the waveguide, transmission measurements were performed on a sample containing only 3 mm-long ridge waveguides, without NbN wires and contact pads. By replacing the probe with another lensed fiber on the second piezo tower, the transmission of a tuneable laser through the entire system including fibers and waveguide was measured. From the measured Fabry-Perot fringes, and particularly from the maximum and minimum transmission (in TE polarisation), $T_{max}$=6.1% and $T_{min}$=1.8%, we deduce that the propagation loss over a 3 mm waveguide length is negligible. Assuming symmetric input/output coupling, and using the standard expression for the Fabry-Perot transmission[22], we derive a coupling efficiency (from fiber input) of 17.4%.

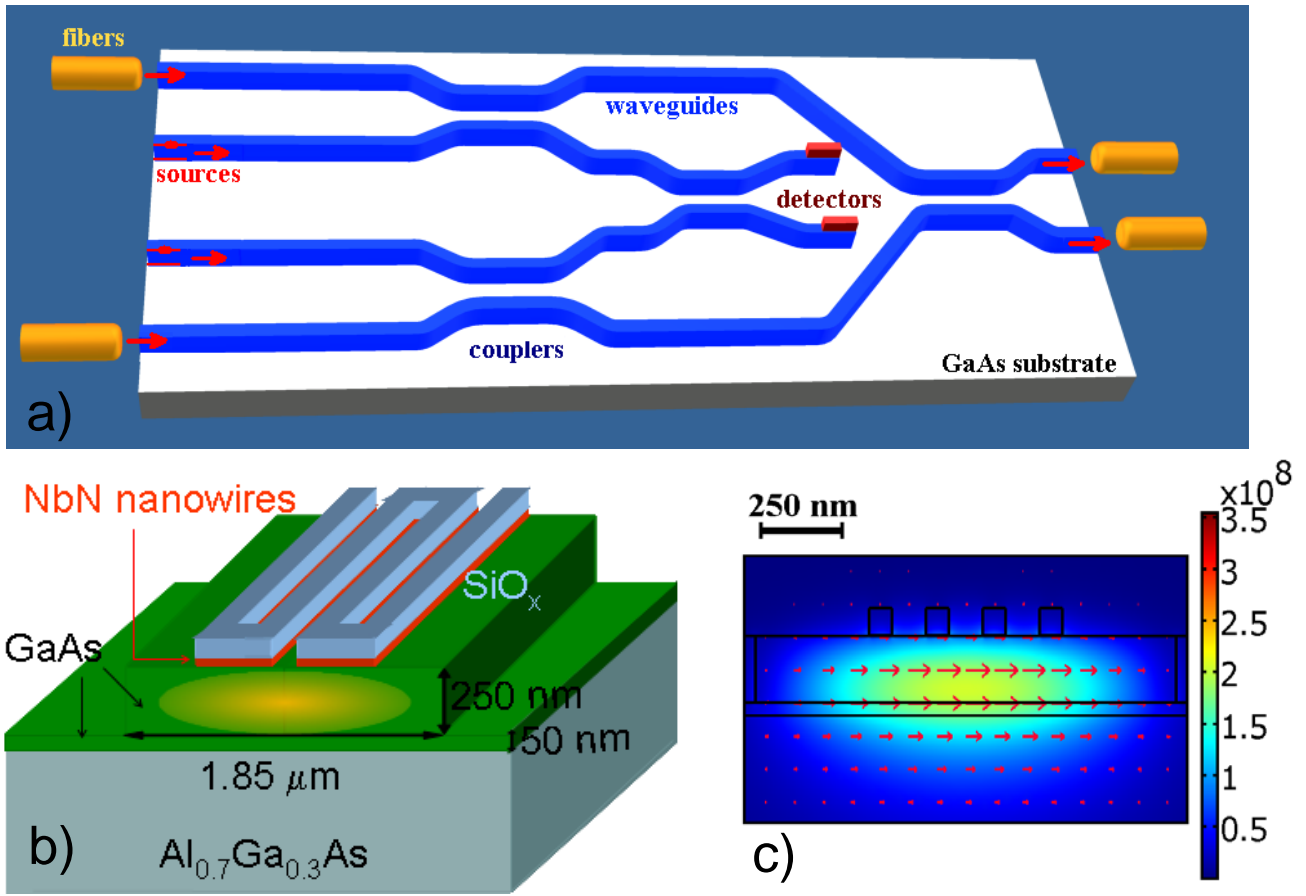

Fig. 1: a) Schematic view of a quantum photonic integrated circuit, which would implement the probabilistic conditional phase shift gate proposed in Ref. [23] by using two internal ancilla photons and two photon-number-resolving detectors; b) WSPD structure; c) Contour and vector plot of the amplitude [V/m] and direction of the electric field for the fundamental mode ($\lambda$=1300 nm) of the WSPD. In the simulation we assume that a 100 nm-thick $SiO_x$ layer is left on top of the wires as a residue of the mask used for patterning the wires.

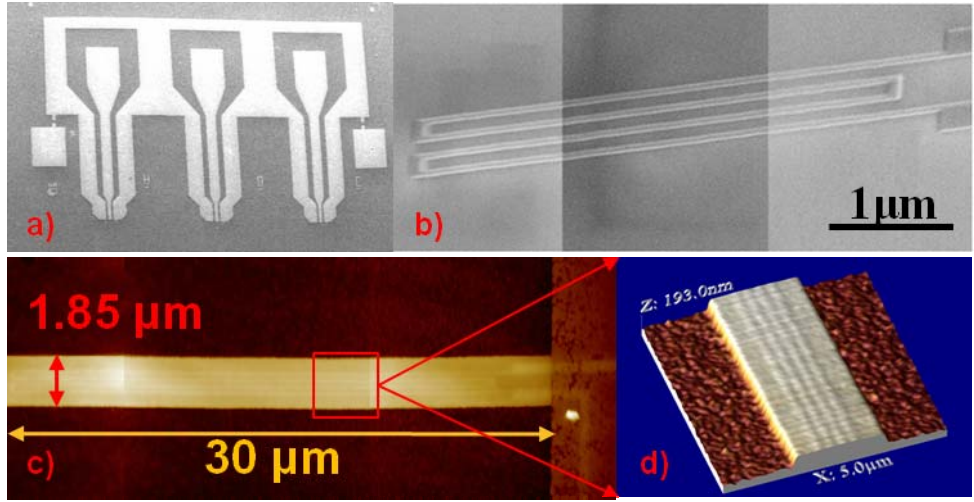

Fig. 2: a) Scanning electron microscope (SEM) micrograph of Ti/Au electrical contacts. b) Collection of three SEM micrographs taken in different regions of a 30μm long WSPD, the nanowires are still covered by the HSQ etching mask; c) Atomic force microscope (AFM) image of the 1.85 μm wide and 30 μm long HSQ mask used for the etching of the waveguide aligned on top of the NbN nanowires; d) AFM Enlarged view of the waveguide HSQ etching mask showing a realignment accuracy better than 100nm.

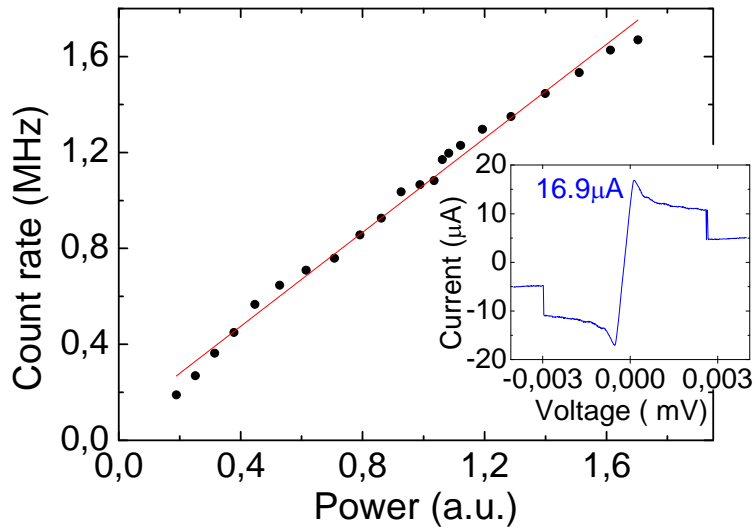

Fig. 3: Count rate as a function of laser power (λ=1300 nm, TE polarisation, $I_b$=9.9 μA), showing a linear behavior and hence operation in the single-photon regime. Inset: IV curve of the WSPD.

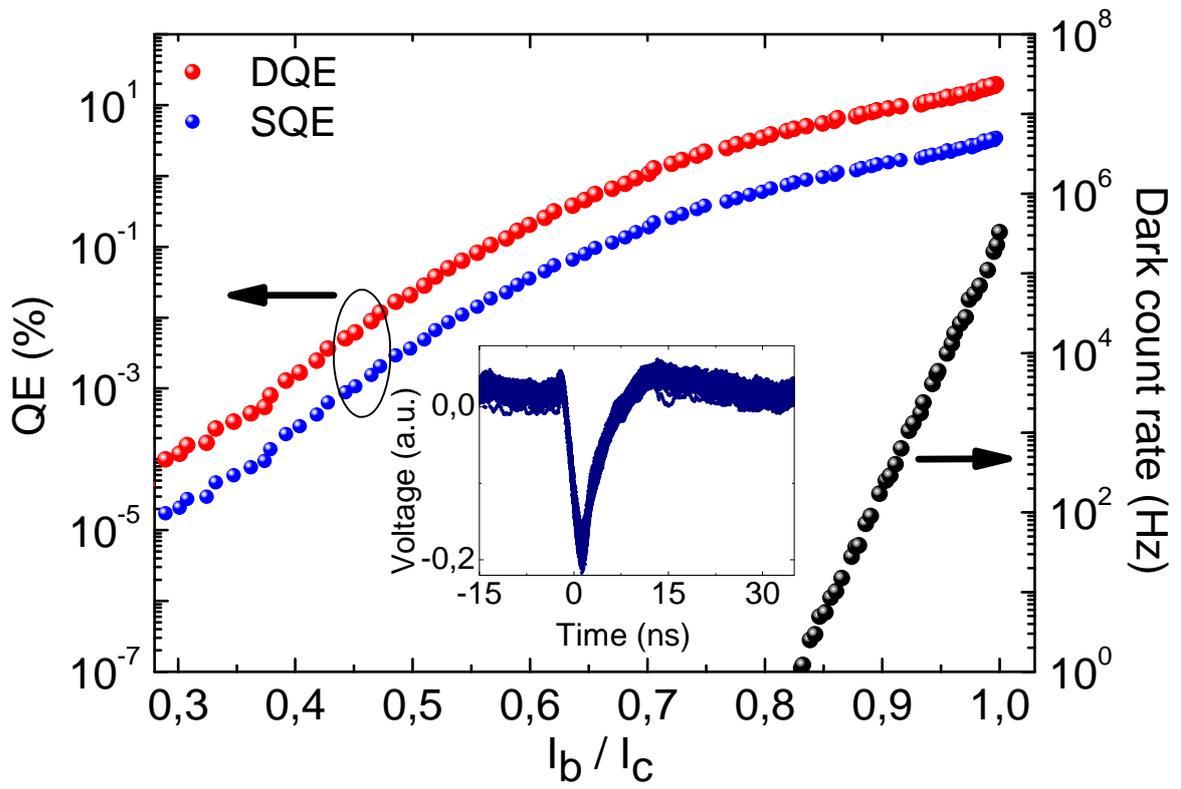

Fig. 4: System QE (blue dots) and device QE (red dots) of a 50 μm-long WSPD under illumination at 1300 nm in the TE polarization (left axis) and dark count rate (black dots, right axis) as a function of the normalized bias current. Inset: WSPD output pulse after 48dB amplification.